\documentclass[12pt]{article}
\usepackage{amsmath,amssymb}
\def\dfrac#1#2{{\displaystyle {#1 \over #2}}}
\begin{document}
\author{{\bf Diana Duplij and Steven Duplij}\thanks{E-mail:
{\tt Steven.A.Duplij@univer.kharkov.ua}} \thanks{Internet:
{\tt http://gluon.physik.uni-kl.de/\~{}duplij}}\\
Kharkov National University, Svoboda Sq. 4, \\
Kharkov 61077, Ukraine}
\title{{\bf
PURINE-PYRIMIDINE SYMMETRY, DETERMINATIVE DEGREE AND DNA
}}
\maketitle
\begin{abstract}
Various symmetries connected with purine-pyrimidine content
of DNA sequences are studied in terms of the intruduced
determinative degree, a new characteristics of nucleotide
which is connected with codon usage. A numerological
explanation of {\bf CG} pressure is proposed. A
classification of DNA sequences is given. Calculations with
real sequences show that purine-pyrimidine symmetry
increases with growing of organization. A new small
parameter which characterizes the purine-pyrimidine symmetry
breaking is proposed for the DNA theory.
\end{abstract}
\newpage
Abstract investigation of the genetic code is a powerful tool in DNA models
construction and understanding of genes organization and expression \cite
{e-sing/berg1}. In this direction the study of symmetries \cite
{fin/fin/mcg,zha1}, application of group theory \cite{hor/hor} and
implication of supersymmetry \cite{bas/tso/jar1} are the most promising and
necessary for further elaboration. In this paper we consider symmetries
connected with purine-pyrimidine content of DNA sequences in terms of the
determinative degree introduced in \cite{e-dup/dup1}.

We denote a triplet of nucleotides by $xyz$, where $x,y,z=\mathbf{C},\mathbf{%
T,A},\mathbf{G}$. Then redundancy means that an amino acid is fully
determined by first two nucleotides $x$ and $y$ independently of third $z$
\cite{e-sing/berg1}. Sixteen possible doublets $xy$ group in 2 octets by
ability of amino acid determination \cite{e-rum1}. Eight doublets have more
``strength'' in sense of the fact that they simply encode amino acid
independently of third bases, other eight (``weak'') doublets for which
third bases determines content of codons. In general, transition from the
``powerful'' octet to the ``weak'' octet can be obtained by the exchange
\cite{e-rum1} $\mathbf{C}\stackrel{\ast }{\Longleftrightarrow }\mathbf{A},\;%
\mathbf{G}\stackrel{\ast }{\Longleftrightarrow }\mathbf{T}$, which we name
``star operation $\left( \ast \right) $'' and call \textit{purine-pyrimidine
inversion}. Thus, if in addition we take into account \textbf{GC} pressure
in evolution \cite{for1} and third place preferences during codon-anticodon
pairing \cite{gra/per/mou}, then 4 nucleotides can be arranged in descending
order in the following way:
\begin{equation}
\begin{array}{cccc}
\text{Pyrimidine} & \text{Purine} & \text{Pyrimidine} & \text{Purine} \\
\mathbf{C} & \mathbf{G} & \mathbf{T} & \mathbf{A} \\
\text{very ``strong''} & \text{``strong''} & \text{``weak''} & \text{very
``weak''}
\end{array}
\label{1}
\end{equation}

Now we introduce a numerical characteristics of the empirical ``strength''
\textemdash\ \textit{determinative degree} $\mathbf{d}_{x}$ of nucleotide $x$
and make transition from qualitative to quanti\-ta\-tive description of
genetic code structure \cite{e-dup/dup1}. It is seen from (\ref{1}) that the
determinative degree of nucleotide can take value $\mathbf{d}_{x}=\mathbf{1},%
\mathbf{2},\mathbf{3},\mathbf{4}$ in correspondence of increasing
``strength''. If we denote determinative degree as upper index for
nucleotide, then four bases (\ref{1}) can be presented as vector-row $\Bbb{V}%
=\left(
\begin{array}{cccc}
\mathbf{C}^{\left( \mathbf{4}\right) } & \mathbf{G}^{\left( \mathbf{3}%
\right) } & \mathbf{T}^{\left( \mathbf{2}\right) } & \mathbf{A}^{\left(
\mathbf{1}\right) }
\end{array}
\right) $. Then the exterior product $\Bbb{M}=\Bbb{V}\times \Bbb{V}$
represents the doublet matrix $\Bbb{M}$ and corresponding rhombic code \cite
{e-kara1}, and the triple exterior product $\Bbb{K}=\Bbb{V}\times \Bbb{V}%
\times \Bbb{V}$ corresponds to the cubic matrix model of the genetic code
which were described in terms of the determinative degree in \cite
{e-dup/dup1}. To calculate the determinative degree of doublets $xy$ we use
the following additivity assumption
\begin{equation}
\mathbf{d}_{xy}=\mathbf{d}_{x}+\mathbf{d}_{y},  \label{dxyz}
\end{equation}
which holds for triplets and for any nucleotide sequence. Then each of 64
elements (codons) of the cubic matrix $\Bbb{K}$ will have a \textit{novel
number characteristics }\textemdash deter\-mi\-na\-tive degree of codon $%
\mathbf{d}_{xyz}=\mathbf{d}_{codon}=\mathbf{d}_{x}+\mathbf{d}_{y}+\mathbf{d}%
_{z}$ which takes value in the range $\mathbf{3\div 12}$. We can also define
the determinative degree of amino acid $\mathbf{d}_{\mathsf{AA}}$ as mean
arithmetic value $\mathbf{d}_{\mathsf{AA}}=\sum \mathbf{d}_{codon}/n_{deg}$,
where $n_{deg}$ is its degeneracy (redundancy). That can allow us to analyze
new abstract amino acid properties in connection with known biological
properties \cite{e-dup/dup1}.

Let us consider a numerical description of an idealized DNA sequence as a
double-helix of two codon strands connected by complementary conditions \cite
{e-sing/berg1}. Each strand is described by four numbers $\left( n_{\mathbf{C%
}},n_{\mathbf{G}},n_{\mathbf{T}},n_{\mathbf{A}}\right) $ and $\left( m_{%
\mathbf{C}},m_{\mathbf{G}},m_{\mathbf{T}},m_{\mathbf{A}}\right) $, where $%
n_{x}$ is a number of nucleotide $x$ in one strand. In terms of $n_{x}$ and $%
m_{x}$ the complementary conditions are
\begin{equation}
n_{\mathbf{C}}=m_{\mathbf{G}},\;m_{\mathbf{C}}=n_{\mathbf{G}},\;n_{\mathbf{T}%
}=m_{\mathbf{A}},\;m_{\mathbf{T}}=n_{\mathbf{A}}.  \label{nm1}
\end{equation}

The Chargaff's rules \cite{e-sing/berg1} for a double-helix DNA sequence
sound as: 1) total quantity of purines and pyrimidines are equal $N_{\mathbf{%
A}}+N_{\mathbf{G}}=N_{\mathbf{C}}+N_{\mathbf{T}}$; 2) total quantity of
adenine and cystosine equal to total quantity of guanine and thymine $N_{%
\mathbf{A}}+N_{\mathbf{C}}=N_{\mathbf{T}}+N_{\mathbf{G}}$; 3) total quantity
of adenine equal to total quantity of thymine $N_{\mathbf{A}}=N_{\mathbf{T}}$
and total quantity of cystosine equal to total quantity of guanine $N_{%
\mathbf{C}}=N_{\mathbf{G}}$; 4) the ratio of guanine and cystosine to
adenine and thymine $v=\left( N_{\mathbf{A}}+N_{\mathbf{T}}\right) /\left(
N_{\mathbf{C}}+N_{\mathbf{G}}\right) $ is approximately constant for each
species. Usually the Chargaff's rules are defined through macroscopic molar
parts which are proportional to absolute number of nucleotides in DNA \cite
{e-sing/berg1}. If we consider a DNA\ double-helix sequence, then $%
N_{x}=n_{x}+m_{x}$. In terms of $n_{x}$ and $m_{x}$ the first three
Chargaff's rules lead to the equations which are obvious identities, if
complimentary (\ref{nm1}) holds. From fourth Chargaff's rule it follows that
the specificity coefficient $v_{nm}$ for two given strands is
\begin{equation}
v_{nm}=\dfrac{n_{A}+m_{\mathbf{A}}+n_{\mathbf{T}}+m_{\mathbf{T}}}{n_{\mathbf{%
C}}+m_{\mathbf{C}}+n_{\mathbf{G}}+m_{\mathbf{G}}}.  \label{vnm}
\end{equation}

The complementary (\ref{nm1}) leads to the equality of coefficients $v$ of
each strand $v_{nm}=v_{n}=v_{m}\equiv v$, and $v$ is connected with \textbf{%
GC} content $p_{\mathbf{CG}}$ in the double-helix DNA as $p_{\mathbf{CG}%
}=1/\left( 1+v\right) $.

We consider another important coefficient: the ratio of purines and
pyrimidines $k$. For two strands from the first Chargaff's rule we obviously
derive $k_{nm}=1$. But for each strand we have
\begin{equation}
k_{n}=\dfrac{n_{\mathbf{G}}+n_{\mathbf{A}}}{n_{\mathbf{C}}+n_{\mathbf{T}}}%
,\;k_{m}=\dfrac{m_{\mathbf{G}}+m_{\mathbf{A}}}{m_{\mathbf{C}}+m_{\mathbf{T}}}
\label{km}
\end{equation}
which satisfy the equation $k_{n}k_{m}=1$ following from complementary.

Let us introduce the determinative degree of \textit{each strand} exploiting
the additivity assumption (\ref{dxyz}) as
\begin{eqnarray}
\mathbf{d}_{n} &=&\mathbf{4}\cdot n_{\mathbf{C}}+\mathbf{3}\cdot n_{\mathbf{G%
}}+\mathbf{2}\cdot n_{\mathbf{T}}+\mathbf{1}\cdot n_{\mathbf{A}},  \label{dn}
\\
\mathbf{d}_{m} &=&\mathbf{4}\cdot m_{\mathbf{C}}+\mathbf{3}\cdot m_{\mathbf{G%
}}+\mathbf{2}\cdot m_{\mathbf{T}}+\mathbf{1}\cdot m_{\mathbf{A}}.  \label{dm}
\end{eqnarray}

The values $\mathbf{d}_{n}$ and $\mathbf{d}_{m}$ can be viewed as
characteristics of the empirical ``strength'' for strands, i.e. ``strand
generalization'' of (\ref{1}). Then we define summing and difference
``strength'' of a double-helix sequence by
\begin{equation}
\mathbf{d}_{+}=\mathbf{d}_{n}+\mathbf{d}_{m},\;\mathbf{d}_{-}=\mathbf{d}_{n}-%
\mathbf{d}_{m}.  \label{d-}
\end{equation}

The first variable $\mathbf{d}_{+}$ can be treated as the summing empirical
``strength'' of DNA (or its fragment). Taking into account the complementary
conditions (\ref{nm1}) we obtain $\mathbf{d}_{+}$ through one strand
variables
\begin{equation}
\mathbf{d}_{+}=\mathbf{7}\cdot \left( n_{\mathbf{C}}+n_{\mathbf{G}}\right) +%
\mathbf{3}\cdot \left( n_{\mathbf{T}}+n_{\mathbf{A}}\right) .  \label{d+n}
\end{equation}

We can also present $\mathbf{d}_{+}$ through macroscopically determined
variables $N_{x}$ as follows $\mathbf{d}_{+}=\mathbf{7}\cdot N_{\mathbf{C}}+%
\mathbf{3}\cdot N_{\mathbf{A}}=\mathbf{7}\cdot N_{\mathbf{G}}+\mathbf{3}%
\cdot N_{\mathbf{T}}$, or through \textbf{GC} and \textbf{AT} contents as $%
\mathbf{d}_{+}=\dfrac{\mathbf{7}}{2}\cdot N_{\mathbf{C+G}}+\dfrac{\mathbf{3}%
}{2}\cdot N_{\mathbf{A+T}}$.

To give sense to the difference $\mathbf{d}_{-}$ we derive
\begin{equation}
\mathbf{d}_{-}=n_{\mathbf{C}}+n_{\mathbf{T}}-n_{\mathbf{G}}-n_{\mathbf{A}}.
\label{d-n}
\end{equation}

We see that the star operation obviously acts as $\left( \mathbf{d}%
_{+}\right) ^{\ast }=\mathbf{d}_{+}$ and $\left( \mathbf{d}_{-}\right)
^{\ast }=-\mathbf{d}_{-}$. From (\ref{d+n})-(\ref{d-n}) it follows the main
statement:

\begin{quotation}
\textit{The biological sense of the determinative degree }$\mathbf{d}$
\textit{is contained in the following purine-pyrimidine relations: }

1) \textit{The sum of the determinative degrees of matrix and complementary
strands in DNA\ (or its fragment) equals to}
\begin{equation}
\mathbf{d}_{+}=\dfrac{\mathbf{7}}{2}\cdot N_{\mathbf{C+G}}+\dfrac{\mathbf{3}%
}{2}\cdot N_{\mathbf{A+T}}.  \label{d73}
\end{equation}

\textit{2) The difference of the determinative degrees between matrix and
complementary strands in DNA\ (or its fragment) exactly equals to the
difference between pyrimidines and purines }\underline{\textit{in one strand}%
}

\begin{equation}
\mathbf{d}_{-}=n_{pyrimidines}-n_{purines},  \label{dpyrpur}
\end{equation}
\textit{where }$n_{pyrimidines}=n_{\mathbf{C}}+n_{\mathbf{T}}$\textit{\ and }%
$n_{purines}=n_{\mathbf{G}}+n_{\mathbf{A}}$\textit{, or it is equal to
difference of purines or pyrimidines between strands}
\begin{equation}
\mathbf{d}_{-}=n_{pyrimidines}-m_{pyrimidines}=m_{purines}-n_{purines}.
\label{dpyrpur1}
\end{equation}
\end{quotation}

We can also find connection between $\mathbf{d}_{+},\mathbf{d}_{-}$ and the
coefficients $k$ and $v$ as follows
\begin{eqnarray}
\mathbf{d}_{+} &=&\dfrac{1}{2}N_{\mathbf{C+G}}\left( 7+3v\right) =N_{\mathbf{%
C+G}}\left( 2+\dfrac{3}{2\cdot p_{\mathbf{CG}}}\right) ,\;  \label{dv} \\
\mathbf{d}_{-} &=&n_{pyrimidines}\left( 1-k_{n}\right) .  \label{dk}
\end{eqnarray}

If we consider one species for which $v=const$ (or $p_{\mathbf{CG}}=const$),
then we observe that $\mathbf{d}_{+}\thicksim N_{\mathbf{C+G}}$, which can
allow us to connect the determinative degree with ''second level'' of
genetic information \cite{for1}. From another side, the ratio $\dfrac{7}{3}$
of coefficients in (\ref{d73}) can play a numerological role in \textbf{CG}
pressure explanations \cite{for1}, and therefore $\mathbf{d}_{+}$ can be
considered as some kind of ``evolutionary strength''.

Now we consider the determinative degree of double-helix sequences in
various extreme cases and classify them. We call a DNA sequence \textit{%
mononucleotide}, \textit{dinucleotide}, \textit{trinucleotide} or \textit{%
full}, if one, two, three or four numbers $n_{x}$ respectively distinct from
zero. Properties of mononucleotide double-helix DNA sequence are in the
Table 1.

\begin{center}
Table 1. Mononucleotide DNA

\begin{tabular}{|c|c|c|c|}
\hline
$n_{x}$ & $\mathbf{d}_{+}$ & $\mathbf{d}_{-}$ & amino acid \\ \hline\hline
$n_{\mathbf{C}}\neq 0$ & $7n_{\mathbf{C}}$ & $n_{\mathbf{C}}$ & \textsf{Pro}
\\ \hline
$n_{\mathbf{G}}\neq 0$ & $7n_{\mathbf{G}}$ & $-n_{\mathbf{G}}$ & \textsf{Gly}
\\ \hline
$n_{\mathbf{T}}\neq 0$ & $3n_{\mathbf{T}}$ & $n_{\mathbf{T}}$ & \textsf{Phe}
\\ \hline
$n_{\mathbf{A}}\neq 0$ & $3n_{\mathbf{A}}$ & $-n_{\mathbf{A}}$ & \textsf{Lis}
\\ \hline
\end{tabular}
\end{center}

The mononucleotide sequences which encode most extended amino acids \textsf{%
Gly} and \textsf{Lis} have negative $\mathbf{d}_{-}$, and the mononucleotide
sequences which encode amino acids \textsf{Pro} and \textsf{Phe} with
similar chemical type of radicals have positive $\mathbf{d}_{-}$.

The dinucleotide double-helix DNA sequences (without mononucleotide parts)
are described in the Table 2.

\begin{center}
Table 2. Dinucleotide DNA

\begin{tabular}{|c|c|c|c|}
\hline
$n_{x}$ & $\mathbf{d}_{+}$ & $\mathbf{d}_{-}$ & amino acid \\ \hline\hline
$n_{\mathbf{C}}\neq 0,n_{\mathbf{G}}\neq 0$ & $7\left( n_{\mathbf{C}}+n_{%
\mathbf{G}}\right) $ & $n_{\mathbf{C}}-n_{\mathbf{G}}$ & \textsf{%
Pro,Arg,Ala,Gly} \\ \hline
$n_{\mathbf{C}}\neq 0,n_{\mathbf{T}}\neq 0$ & $7n_{\mathbf{C}}+3n_{\mathbf{T}%
}$ & $n_{\mathbf{C}}+n_{\mathbf{T}}$ & \textsf{Pro,Phe,Leu,Ser} \\ \hline
$n_{\mathbf{C}}\neq 0,n_{\mathbf{A}}\neq 0$ & $7n_{\mathbf{C}}+3n_{\mathbf{A}%
}$ & $n_{\mathbf{C}}-n_{\mathbf{A}}$ & \textsf{Pro,Gly,Asn,Tur,His} \\ \hline
$n_{\mathbf{G}}\neq 0,n_{\mathbf{T}}\neq 0$ & $7n_{\mathbf{G}}+3n_{\mathbf{T}%
}$ & $n_{\mathbf{T}}-n_{\mathbf{G}}$ & \textsf{Gly,Leu,Val,Cys,Trp} \\ \hline
$n_{\mathbf{G}}\neq 0,n_{\mathbf{A}}\neq 0$ & $7n_{\mathbf{G}}+3n_{\mathbf{A}%
}$ & $-n_{\mathbf{G}}-n_{\mathbf{A}}$ & \textsf{Gly,Glu,Arg,Lys} \\ \hline
$n_{\mathbf{T}}\neq 0,n_{\mathbf{A}}\neq 0$ & $3\left( n_{\mathbf{T}}+n_{%
\mathbf{A}}\right) $ & $n_{\mathbf{T}}-n_{\mathbf{A}}$ & \textsf{%
Leu,Asn,Tur,TERM} \\ \hline
\end{tabular}
\end{center}

The trinucleotide DNA can be listed in the similar, but more cumbersome way.
The full DNA sequences consist of nucleotides of all four types and
described by (\ref{d+n})-(\ref{d-n}).

The introduction of the determinative degree allows us to single out a kind
of double-helix DNA sequences which have an additional symmetry. We call a
double-helix sequence \textit{purine-pyrimidine symmetric}, if
\begin{equation}
\mathbf{d}_{-}=0,  \label{d0}
\end{equation}
i.e. its empiric ``strength'' vanishes. From (\ref{d-n}) it follows
\begin{equation}
n_{\mathbf{C}}+n_{\mathbf{T}}=n_{\mathbf{G}}+n_{\mathbf{A}},  \label{n}
\end{equation}
i.e. $k_{n}=k_{m}=1$, which can be rewritten for one strand
\begin{equation}
n_{pyrimidines}=n_{purines}  \label{nn}
\end{equation}
or as equality of purines and pyrimidines in two strands
\begin{eqnarray}
n_{pyrimidines} &=&m_{pyrimidines},  \label{nm} \\
n_{purines} &=&m_{purines}.  \label{nm2}
\end{eqnarray}

The purine-pyrimidine symmetry (\ref{n}) has two particular cases:
\begin{eqnarray}
&&1)\;
\begin{array}{c}
n_{\mathbf{C}}=n_{\mathbf{G}}, \\
n_{\mathbf{T}}=n_{\mathbf{A}},
\end{array}
-\text{symmetric DNA,}  \label{ns1} \\
&&2)\;
\begin{array}{c}
n_{\mathbf{C}}=n_{\mathbf{A}}, \\
n_{\mathbf{T}}=n_{\mathbf{G}},
\end{array}
-\text{antisymmetric DNA.}  \label{ns2}
\end{eqnarray}

The first case corresponds to the Chargaff's rule applied to a single strand
which approximately holds for long sequences \cite{for2}, and so it would be
interesting to compare transcription and expression properties of symmetric
and antisymmetric double-helix sequences.

We have made a preliminary analysis of real sequences of several species
taken from GenBank (2000) in terms of the determinative degree. It were
considered 10 complete sequences of \textit{E.coli} (several genes and full
genomic DNA 9-12 min.), 12 complete sequences of \textit{Drosophila
melanogaster} (crc genes), 10 complete sequences of \textit{Homo sapiens}
Chromosome 22 (various clones), 10 complete sequences of \textit{Homo sapiens%
} Chromosome 3 (various clones). We calculated the nucleotide content $N_{%
\mathbf{C}},N_{\mathbf{T}},N_{\mathbf{G}},N_{\mathbf{A}}$ and the
determinative degree characteristics $\mathbf{d}_{+},\mathbf{d}_{-},q=%
\mathbf{d}_{-}/\mathbf{d}_{+},k_{n}$ and $v$ for every sequence. Then we
averaged their values for each species. The result is presented in the Table
3.

\begin{center}
Table 3. Mean determinative degree characteristics of real sequences

\begin{tabular}{|c|c|c|c|c|c|}
\hline
sequence & $\dfrac{1}{n}\sum \mathbf{d}_{+}$ & $\dfrac{1}{n}\sum \mathbf{d}%
_{-}$ & $\dfrac{1}{n}\sum q\cdot 10^{3}$ & $\dfrac{1}{n}\sum k_{n}$ & $%
\dfrac{1}{n}\sum v$ \\[5pt] \hline\hline
\textit{E.coli} & 90806 & -138 & -6.8 & 1.07 & 1.38 \\ \hline
\textit{Drosophila} & 7325 & -70 & -8.9 & 1.09 & 1.31 \\ \hline
\textit{Homo sap.} Chr.22 & 337974 & 6865 & 1.46 & 0.987 & 1.14 \\ \hline
\textit{Homo sap.} Chr.3 & 806435 & -1794 & -2.29 & 1.021 & 1.55 \\ \hline
\end{tabular}
\end{center}

First of all we observe that all real sequences have high purine-pyrimidine
symmetry (smallness of parameter $q$). Also we see that the relation of
purines and pyrimidines in one DNA strand $k_{n}$ is very close to unity,
therefore we have a new small parameter in the DNA theory $\left(
k_{n}-1\right) $ (or $q$), which characterizes the purine-pyrimidine
symmetry breaking. This can open possibility for various approximate and
perturbative methods application. Second, we notice from Table 3 that the
purine-pyrimidine symmetry increases in direction from protozoa to mammalia
and is maximal for human chromosome. It would be worthwhile to provide a
thorough study of purine-pyrimidine symmetry and codon usage in terms of the
introduced determinative degree by statistical methods, which will be done
elsewhere.

\textbf{Acknowledgments}. Authors would like to thank G. Shepelev for
providing with computer programs, S. Gatash, V. Maleev and O. Tretyakov for
fruitful discussions and J. Bashford, G. Findley and P. Jarvis for useful
correspondence and reprints.
%\newpage

\end{document}